\documentclass[conference, 10pt, a4paper]{IEEEtran}
\IEEEoverridecommandlockouts
\usepackage{graphics} 
\usepackage{epsfig} 
\usepackage{amsmath} 
\usepackage{amssymb}  
\usepackage{color}  
\usepackage{epstopdf}

\usepackage{subfigure}
\usepackage[square, comma, sort&compress, numbers]{natbib}

\ifCLASSINFOpdf
\else
\fi

\begin{document}
\setlength{\parskip}{0pt}
\setlength{\itemsep}{0pt}
\setlength{\parsep}{0pt}
\setlength{\partopsep}{0pt}
\title{Topic Model Based Behaviour Modeling and Clustering Analysis for Wireless Network Users}
\author{
\IEEEauthorblockN{Bingjie Leng, Jingchu Liu, Huimin Pan, Sheng Zhou, and Zhisheng Niu}
\IEEEauthorblockA{Tsinghua National Laboratory for Information Science and Technology \\ Department of Electronic Engineering\\Tsinghua University,
Beijing 100084, China\\
Email:  \{lengbj14, liu-jc12, phm13\}@mails.tsinghua.edu.cn, \{sheng.zhou, niuzhs\}@tsinghua.edu.cn
}
}

%

\maketitle

\begin{abstract}
User behaviour analysis based on traffic log in wireless networks can be beneficial to many fields in real life: not only for commercial purposes, but also for improving network service quality and social management. We cluster users into groups marked by the most frequently visited websites to find their preferences. In this paper, we propose a user behaviour model based on \emph{Topic Model} from document classification problems. We use the logarithmic TF-IDF (term frequency - inverse document frequency) weighing to form a high-dimensional sparse feature matrix. Then we apply LSA (Latent semantic analysis) to deduce the latent topic distribution and generate a low-dimensional dense feature matrix. K-means++, which is a classic clustering algorithm, is then applied to the dense feature matrix and several interpretable user clusters are found. Moreover, by combining the clustering results with additional demographical information, including age, gender, and financial information, we are able to uncover more realistic implications from the clustering results.
\end{abstract}

\bigskip

%
\IEEEpeerreviewmaketitle

\section{Introduction}
Thanks to the wide adoption of smart devices such as smart phones and tablets, nowadays people can perform an unprecedented number of tasks online, ranging from news and finance to social and gaming. As a consequence, Internet browsing log in wireless networks has become an essential source of information for analyzing users' hidden preferences and inferring their real life behaviour.  With a deeper understanding on the usage pattern of mobile users, network service providers are able to provide more personalized services and improve the service quality as well. Users' browsing interests are also helpful in fields such as urban planning, mobile advertisement, transportation, education, etc \cite{lin2015clusoaf, gonzalez2008understanding, bicocchi2014investigating}.

The most naive way to extract user behaviour from the Internet browsing dataset is to observe the long-term global statistics of various websites. But in this situation, individual differences will be covered up. On the contrary, if we focus on the analysis on one single user, the similarity between users' browsing habits will be ignored. Hence, clustering becomes an efficient method to strike a balance between these two extremes and extract the average behaviour of a group of users who have similar browsing history. Therefore, we design and implement a process to cluster similar users into groups, each of which is labeled by the type of frequently visited websites.

To apply clustering algorithms, the first step is to represent users with a profile vector through user behaviour modeling. In this paper, we propose a user behaviour modeling method based on the topic model, which is originally proposed for document classification, to generate an original profile matrix. To enhance the discriminative power of the original matrix, we apply TF-IDF (term frequency - inverse document frequency) weights to regenerate a feature matrix with large dimensionality. With methods in Latent semantic analysis (LSA) \cite{dumais2004latent}, we are able to get a low-dimensional feature matrix reflecting the distribution of different topics of all the users. Finally, clustering algorithms such as K-means++ can be applied to the final feature matrix and the clustering results are analyzed.

Concretely, we make the following contributions in this paper:
\begin{itemize}
\item We analyze the similarity and differences between network user modeling and document classification, and propose to utilize text mining algorithms for network user modeling problems.
\item Based on the analysis on our dataset, we utilize logarithmic TF-IDF to generate sparse feature matrix and use LSA for topic discovery and dimensionality reduction. To our knowledge, this is the first study to analyze user behaviour with a combination of these tools.
\item We extract users' interests by clustering users with similar browsing habits into groups. We also examine our clustering results with additional demographical information including age, gender, and financial information on the campus during five months. Obvious preference differences are found between different genders and age. It helps us explain our clustering findings accordingly and proves that our algorithm can work effectively. Moreover, our findings can help with campus management in many aspects.
\end{itemize}

The rest of the paper is outlined as follows. Section \ref{sec_relateed_work} introduces related work about user behaviour analysis in WLAN. Section \ref{sec_user_modeling} presents the network user behavior modeling problem and explain its analogy with topic modeling in document classification. In Section \ref{sec_implementation}, we introduce our datasets and the details of our algorithm implementation. In Section \ref{sec_experimental results}, we present the clustering results and explain the findings. Finally, in Section \ref{sec_conclusion}, we conclude and discuss future work.

\begin{table*}[!t]
\caption{Analogy between User Behaviour Modeling and Topic Modeling in Document Classification}
\label{table}
\centering
\begin{tabular}{|c||c|}
\hline
\textbf{Elements in Network User Behavior Modeling} & \textbf{Elements in Document Classification}\\
\hline
User & Document\\
\hline
Tuple of Domains/ Locations/ Time & Words\\
\hline
Value for Tuple(Total Bytes/ Duration/ Request Numbers) & Frequency of Words\\
\hline
User Clustering & Document Classification\\
\hline
\end{tabular}
\end{table*}
\section{Related Work}
\label{sec_relateed_work}
With the rapid development of wireless networks, the potential of user behaviour analysis has brought up tremendous attention recently.
The most common method for user clustering is by applying K-means on raw profile matrix. For example, the web browsing similarity among users of a university laboratory is examined using web log data in in \cite{xu2010web}. Besides, in \cite{karamolegkos2007user}, the authors evaluate the effectiveness of K-Means and spectral clustering and discuss the effect of the profile size to the clustering results. The authors in \cite{paireekreng2009client} discuss this personalization problem from another perspective. They apply K-Means to construct user profiles using device-side traffic logs. The authors also point out that the extra information on gender may help with the problem in future work.

Besides K-means, other clustering algorithms are also investigated in the literature. The authors in \cite{xie2001web} present a Greedy Clustering Using Belief Function
(GCB) algorithm. The representatives of the clusters are picked iteratively, so that the current representative is well separated from the former ones. In \cite{moghaddam2011multidimensional}, the authors analyze the online activity and mobility for thousands of mobile users. They use Self-Organized Map (SOM) for discovering mobile users' trends. In \cite{moghaddam2010data}, the authors use a information theoretic co-clustering method on user-domain matrix and consider the effect of location.

Many similarity measures between users are studied. The authors in \cite{elsherief2014quest} use dataset containing traces of 34 users and mainly deal with temporal dimension. In their work, different similarity matrices including Cosine Similarity, Hellinger distance and SVD are compared with each other. The authors in \cite{xie2001web} apply a belief function as the similarity measure.

A closely recent related work is \cite{giri2014user}. Same as us, the authors study user behaviour modeling in a perspective of topic modeling. By way of analogy, the authors find that there are many similarities between user behaviour profiling and document classification in natural language processing, which is also used in our modeling. But in their work, the authors generate their feature matrix by oversampling the URLs, while we use TF-IDF weights. The data comes from cellualr networks in their paper. But the datasets we use are clollected from WLAN. In \cite{sundin2012word}, the author proposes a similar procedure in analyzing user preferences with our algorithm for webpage prefetching. But the data set used in the paper is small and there are only around 100 domains included. Therefore, user behaviour cannot be fully presented.

\section{User Behaviour Modeling}
\label{sec_user_modeling}
There are many similarities between user behaviour analysis and document classification.
We will introduce our user behaviour modeling by comparing the nature of user behaviour profiling with the properties of a document in this section.

The main purpose of document classification is to find the hidden topics of the corpus of documents by unsupervised learning algorithms. In Table \ref{table}, the correspondences between user behaviour modeling and topic modeling are listed. In user behaviour modeling, we plan to cluster users with similar internet browsing records, which is similar as clustering documents with similar topics. The content of the internet browsing records can be considered as words existing in a document to some degree, although the statistic properties of internet records are more complicated than term frequency in a document. These findings encourage us to use tools in document classification to classify users' internet browsing preferences.

Due to this similarity, tools designed for document classification may also be fit to network user behaviour modeling. In this work, we use TF-IDF statistics and LSA to cluster users with similar network usage behavior and preferences. TF-IDF is a nonlinear weighting method to reveal the correlation between words and topics. LSA is an efficient technique to mine topics and reduce complexity at the same time.

\section{Datasets and Clustering Implementation}
\label{sec_implementation}
In this section, we introduce the properties of the campus WiFi dataset used in this paper. We also outline the implementation of our algorithm to extract user behaviour, with detailed explanation for each algorithm step. The steps include data pre-processing, user profile matrix generalization, generating the feature matrix with TF-IDF, LSA and clustering.

\subsection{Overview of Datasets}
In this paper, we use the Internet browsing data extracted from a WiFi network in a Chinese university campus. This dataset has been collected from September, 2014 to January, 2015, and contains the browsing records of over 28,000 anonymous users.

User's browsing activity is aggregated into network sessions, which are defined to be the period in which users utilize the network with pauses that are less than 5 minutes. If a user stops exchanging information with the network for more than 5 minutes, the session ends and activities from that moment on are recorded in the following session. Each session contains the information of network browsing activity of a user, including user ID, session starting time, session duration, location, first-level or second-level domains, ISP names, HTTP request number and predefined service classes.

Except for the internet browsing dataset, a complementary dataset containing users' personal profile is also given. Each row in this dataset represents for one user's profile, including user ID, gender, the year of birth, the year of enrolment and type of degree.
Besides, we are also provided with a dataset containing all trading information of users' campus cards during the same period of time.
These additional information gives us an opportunity to interpret the results of user behaviour clustering and learn more about users' behaviour.

\subsection{Data Pre-processing and Overall Statistics}
\label{sec_preprocessing}

To facilitate algorithm processing, we extract the network browsing information of each user from the dataset and form a user profile matrix. Each row represents one user and each column represents the values of one profile dimension. Based on the types of information available in the original dataset, there are several ways for us to formulate the user profile, including the domain names, time of day or week, locations or encapsulating a multiply of these informations. In this work, we only use the domain names of the websites that the users access over the WiFi.

As for the values in the feature matrix, four statistics of user behaviour are available, including the total communication bytes, the aggregated duration, the cumulative HTTP request number, and the cumulative number of sessions of one user on a certain domain during these five months. In this paper, we choose the total communication bytes as the profile matrix entry value.

After pre-processing, we find that there are 893 first-level domains covered in the dataset. We investigate the basic statistics of the resulting profile matrix. We find that the medians of rows (or equivalently, users) are all zeros, which indicates that all the users only visit a highly limited range of domains. Besides, only 5\% of the domains have non-zero median browsing activity among users. This indicates that most of these domains are only accessed by a small amount of users and users' domain preferences are quite centralized.
We rank the domains with respect to their median browsing intensity and show the histogram of browsing intensity of three of the top-10 domains in Fig. 1.
As can be seen, most of the users would visit these domains with high traffic. The top domains which are visited most include \textit{baidu.com}, \textit{qq.com}, \textit{sinaimg.cn}, etc.

\begin{figure}[!t]
\centering
\includegraphics[width=3.5in]{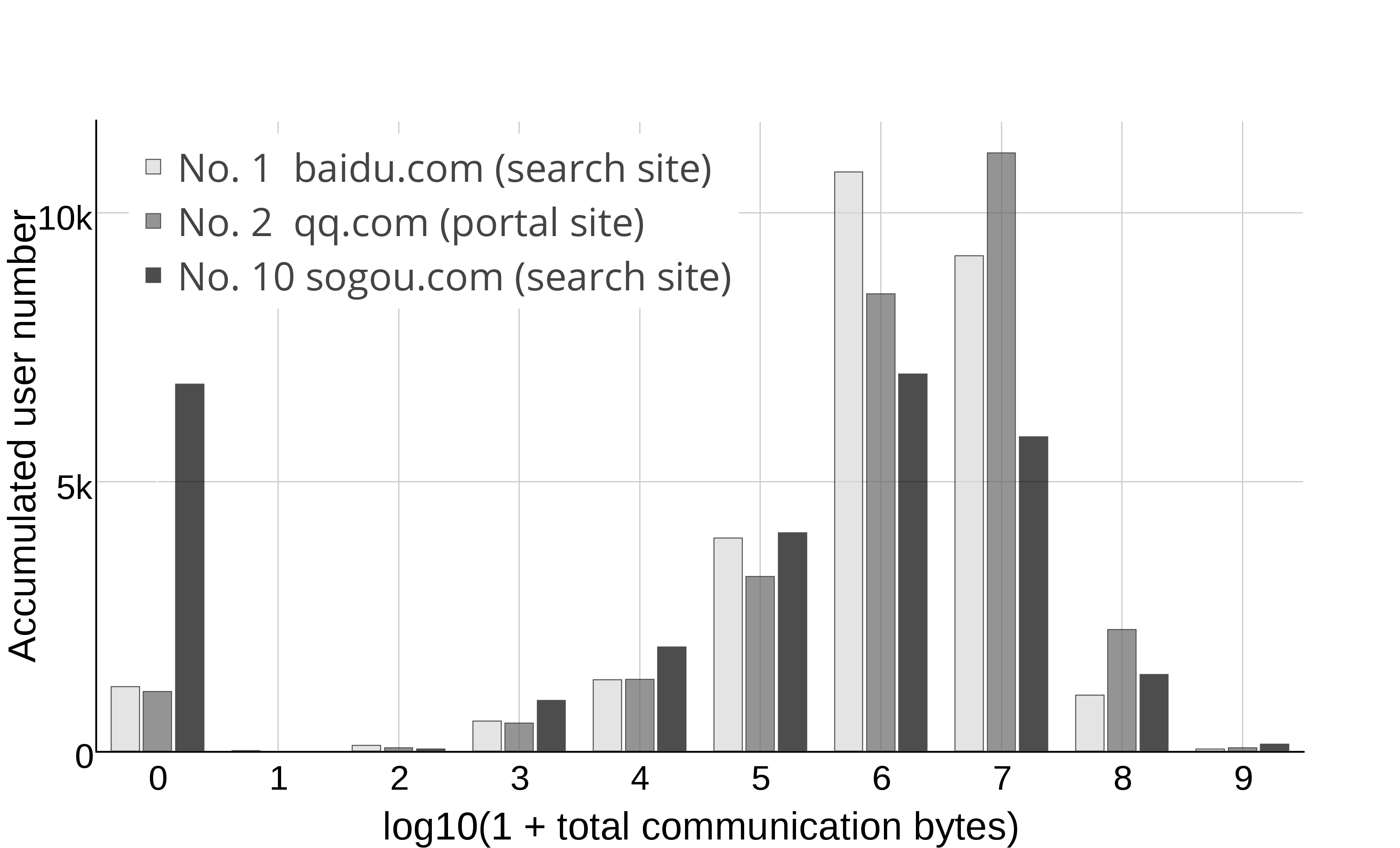}
\caption{Histogram of total bytes for user browsing.}
\label{fig_median}
\end{figure}

\subsection{Generation of Feature Matrix With TF-IDF}
The matrix generated in the last section needs to be normalized before being used for clustering. But with simple normalization of each row, the weights in the matrix (i.e. TF) cannot represent the importance of various domains to one user precisely. This is because there are some domains that are almost visited by everyone, such as search sites like \textit{baidu.com} and portal sites like \textit{qq.com}. It is obvious that their normalized weights for most users would dominate over other domains and user preferences would be biased toward these dominating domains. For this reason, we use TF-IDF as a method to reduce the weights of this kind of popular domains and generate a new feature matrix which will be used in the following steps.

TF-IDF is a non-linear transformation which is widely used in document classification to unveil the true discriminative power of a certain term to some document. It is the product of term frequency (TF) and inverse document frequency (IDF). In document classification tasks, TF means the frequency for a word to appear in a document. We extend this concept to user clustering with network traffic log. Specially, TF-IDF is now corresponded to the ratio between the total bytes consumed on each domain to the total bytes consumed on all the domains that appear in the dataset. Still, the traditional form of TF cannot be directly applied to the network traffic log because of the large dynamic range across the total communication bytes of different domains: the largest and the smallest can be different to 8 orders of magnitude. To cope with this problem, we apply logarithmic operation on TF terms to narrow down the dynamic range. Concretely, assume that there are $N_u$ users and $N_d$ domains involved. Let $B_{ij}$ denote the total bytes of the $j$th domain for the $i$th user, where $j \in \{1, 2,..., N_d\}$ and $i \in \{1, 2,..., N_u\}$. The TF weight for the $j$th domain and user $i$, i.e. $\mathrm{TF}_{ij}$ can be expressed as:
\begin{equation}
\label{eq_tf}
\mathrm{TF}_{ij} = 1 + \log(\frac{B_{ij}}{\sum_{j=1}^{N_d}{B_{ij}}}).
\end{equation}

The inverse document frequency (IDF) is introduced to show how much information a domain provides. IDF can be expressed as:
\begin{equation}
\label{eq_idf}
\mathrm{IDF}_{ij} = \log(\frac{N_u}{n_j}),
\end{equation}
where $n_j$ denotes the number of users ever visited the $j$th domain.

Using the above notations, we can get the new feature matrix $F$, in which each element is given by:
\begin{equation}
\label{eq_tfidf}
F_{ij} = \mathrm{TF}_{ij} \cdot \mathrm{IDF}_{ij}.
\end{equation}


\subsection{Dimensionality Reduction Using Truncated SVD}
To analyze user preferences by clustering, the distances between feature vectors should be computed. But the large (row) dimension of the user feature matrix may render common distance measures, e.g. cosine distance, useless due to the so called "curse of dimensionality". Besides, the large dimensionality also makes clustering algorithms slow. To solve this problem, we assume network activity to follow a topic model and apply LSA to reduce the dimensionality of the sparse feature matrix and get a dense one with much smaller number of dimensions.

Topic modeling assumes the following generative process for text documents: each document can be expressed as a distribution over a collection of $T$ topics $\{p_t, t = 1, 2... T\}$, and each topic is manifested through a probability distribution over a collection of $W$ words $\{p_{w|t}, w = 1,2,3...,W\}$. The overall probability of a word $w$ to appear in a document is $p_w=\sum_{t=1}^{T}{p_{w|t}p_t}$. In user behaviour analysis, the traffic of each user on domains can be assumed to follow a similar generative process.

A common way to recover the hidden parameters of such latent distribution is LSA.  Specially, we can factorize the feature matrix $F$ using SVD. The decomposition of $F$ is given by:
\begin{equation}
\label{eq_SVD}
F = U \Sigma V^T,
\end{equation}
where $U$ and $V$ are orthogonal matrices and $\Sigma$ is a diagonal matrix. If we further assume the domain corpus contians only a small number of topics, we can perform the truncated SVD to extract part of the singular vectors and values. The approximation of $F$ can be expressed as follows:
\begin{equation}
\label{eq_SVD_k}
F_M = U_M \Sigma_M V_M^T,
\end{equation}
where the M largest singular values in $\Sigma$ and their corresponding singular vectors from U and V are selected. $U_M$ can be treated as the final feature matrix used for clustering, which is of much lower dimensionality than the original sparse feature matrix. In the meantime, observations would be extracted from $V_M$, which contains information about the relationship between different domains.

\subsection{K-Means Clustering}
As for the clustering algorithm, we finish the user clustering based on the classic K-Means algorithm. The K-Means algorithm divides all the users into $K$ clusters and guarantees that the within-cluster sum of squared distances, a.k.a. \textit{inertia}, is minimized.  Due to the randomness of initial seeds selection in standard K-Means algorithm, we use K-Means++ \cite{arthur2007k} instead, which performs multiple K-Means runs with different random initialization and selects the best clustering results.

\section{Evaluation of Implementation and Experimental Results}
\label{sec_experimental results}
In this section, we first evaluate and choose the parameters of our implementation based on numerical analysis. Then, we describe the experimental results and give corresponding advice on campus management.

\subsection{Evaluation of Implementation}
\begin{figure}[!t]
\centering
\includegraphics[width=3in]{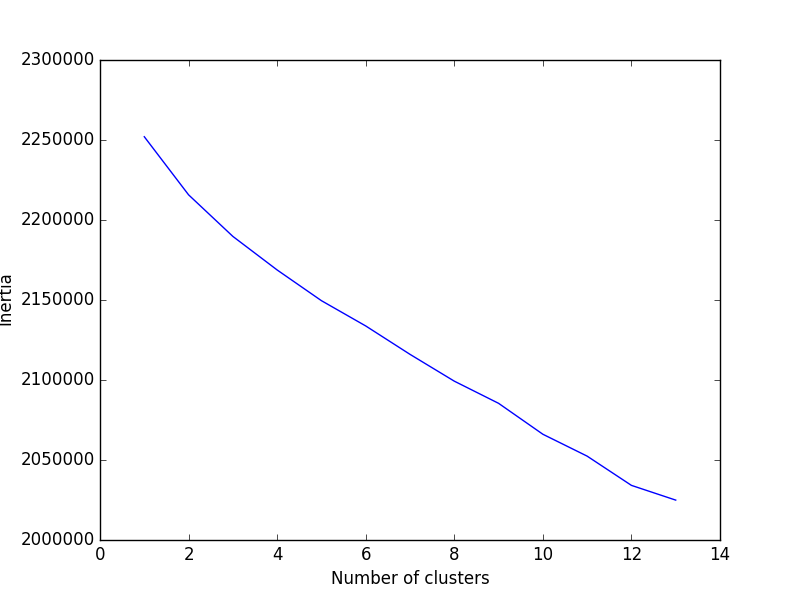}
\caption{The trend of inertia changing with the number of clusters $K$.}
\label{fig_ksweep}
\end{figure}

The selection of the cluster number $K$ is an essential issue, as this may influence the clustering results a lot. Fig. \ref{fig_ksweep} shows the values of inertia when different values of $K$ are applied, from 1 to 13. As we can see, the value of inertia goes down with the increase of $K$.

Except for $K$, the number of the selected singular values $M$ in LSA must be decided as well. In Table \ref{table_M}, the running time of the whole implementation process with different values of $M$ ia given. As $M$ goes up, the complexity of the algorithm increases, so that the running time increases. But the smaller $M$ is, the more information of the original feature matrix will be ignored. Therefore, a modest $M$ should be selected because of the tradeoff between complexity and information integrity.
\begin{table}
\caption{Running time of the whole process}
\label{table_M}
\centering
\begin{tabular}{|c||c|}
\hline
$\textbf{M}$ & \textbf{Running Time (s)}\\
\hline
100 & 11.66\\
\hline
200 & 22.37\\
\hline
300 & 30.13\\
\hline
400 & 43.06\\
\hline
500 & 56.82\\
\hline
600 & 84.07\\
\hline
700 & 93.15\\
\hline
800 & 112.17\\
\hline
\end{tabular}
\end{table}

\subsection{Experimental Results}
Based on the analysis above, we select the top 80 singular components in LSA, i.e., $M=80$, to reduce the computation complexity and get enough information for clustering at the same time. Since different numbers of clusters $K$ may incur different clustering results, $K$ is chosen to be $8$ as an example to explain the clustering results in the following.

\begin{figure}[!t]
\centering
\includegraphics[width=3.5in]{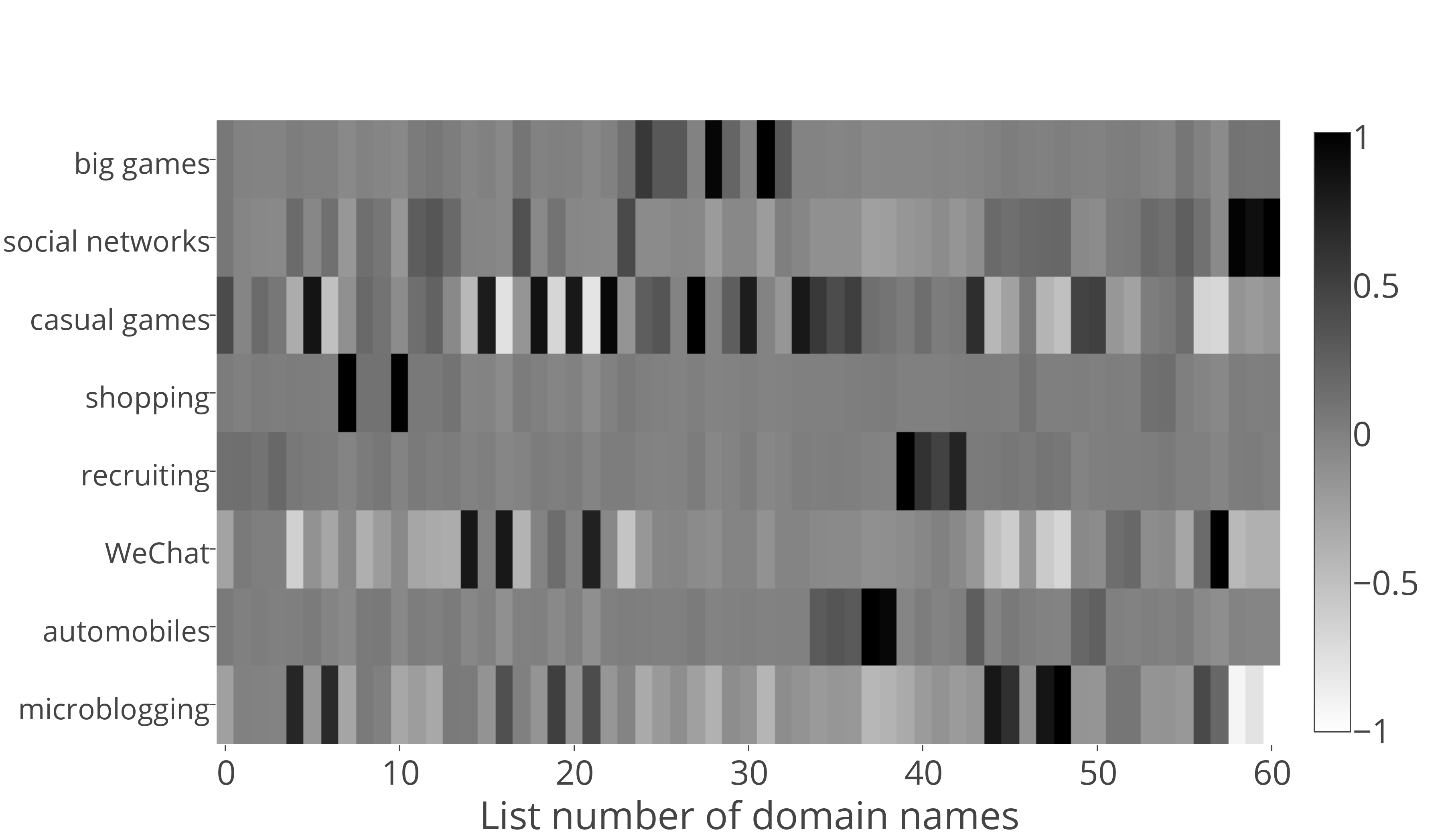}
\caption{TF-IDF distribution of domains in each cluster.}
\label{fig_domain_cluster}
\end{figure}

The average TF-IDF distribution of domains in each user cluster is shown in Fig. \ref{fig_domain_cluster}. The horizontal axis lists 61 domain names which include all the top-10 domains with the largest TF-IDF weights of each cluster. Each cluster is marked on the Y-axis with the summarized topic or by the domain name with the largest TF-IDF weight within this cluster.
As we can see, there are several hot domains in each cluster, which indicates the preferences of users within this cluster. The clusters include preferences for automobiles, games, social networks, online shopping websites and so on. To be clearer, we list three typical domain names in each cluster with the largest TF-IDF weights and the topic of each cluster is summarized based on these domains in Table \ref{table_cluster}. For clusters with no obvious topic, but only preferring some general websites such as WeChat and Baidu, the topic of these clusters are the services correspoding to the largest TF-IDF weight in each of them.

\begin{table*}[!t]
\caption{List of summarized topics and top three domains in each cluster with TF-IDF weighting.}
\label{table_cluster}
\centering
\begin{tabular}{|c||c|}
\hline
\textbf{Topics} & \textbf{Top Three Domains}\\
\hline
Big Games & duowan.com (game news), yy.com (game webcast), battlenet.com.cn \\
\hline
Social Networks & xnimg.com, renren.com, xnpic.com (all related to renren.com)\\
\hline
Casual Games & 66game.com.cn, 7.qq.com (webgame), zg.qq.com (webgame)\\
\hline
Shopping & geilicdn.com (mobile shopping), koudai.com (mobile shopping), meitu.com\\
\hline
Recruiting & yingjiesheng.com, 51job.com, zhaopin.com (recruitment sites)\\
\hline
WeChat & weixin.qq.com (WeChat), apple.com, qq.com\\
\hline
Automobiles & autohome.com.cn, auto.suhu.com, auto.qq.com\\
\hline
Microblogging & sinaimg.cn, weibo.cn, blog.sina.cn (mobile blog)\\
\hline
\end{tabular}
\end{table*}

As a comparison with our implementation, we list the top three domains in each cluster without TF-IDF weighting in Table \ref{table_cluster_notfidf}. The values in the input matrix of LSA here are the simple row normalization results of the original feature matrix generated in Section \ref{sec_preprocessing}. From the clustering results, we find that all the top domains are some general websites within multiple service classes. Unlike Table \ref{table}, it is hard to define a particular topic or interest for each cluster, which indicates the process with TF-IDF raises the effectiveness of clustering a lot.

\begin{table}[!t]
\caption{List of top three domains in each cluster without TF-IDF weighting.}
\label{table_cluster_notfidf}
\centering
\begin{tabular}{|c||c|}
\hline
\textbf{Cluster No.} & \textbf{Top Three Domains}\\
\hline
1 & sinaimg.cn, sina.cn, sinastorage.com \\
\hline
2 & qq.com, gtimg.com, speed.qq.com\\
\hline
3 & baidu.com, bdimg.com, baidu-img.cn\\
\hline
4 & xnimg.com, renren.com, xnpic.com\\
\hline
5 & apple.com, verisign.com, autonavi.com\\
\hline
6 & weixin.qq.com, apple.com, qq.com\\
\hline
7 & 360.cn, microsoft.com, mi.com\\
\hline
8 & 163.com, umeng.com, youdao.com\\
\hline
\end{tabular}
\end{table}

\begin{figure}[!t]
\centering
\includegraphics[width=3.5in]{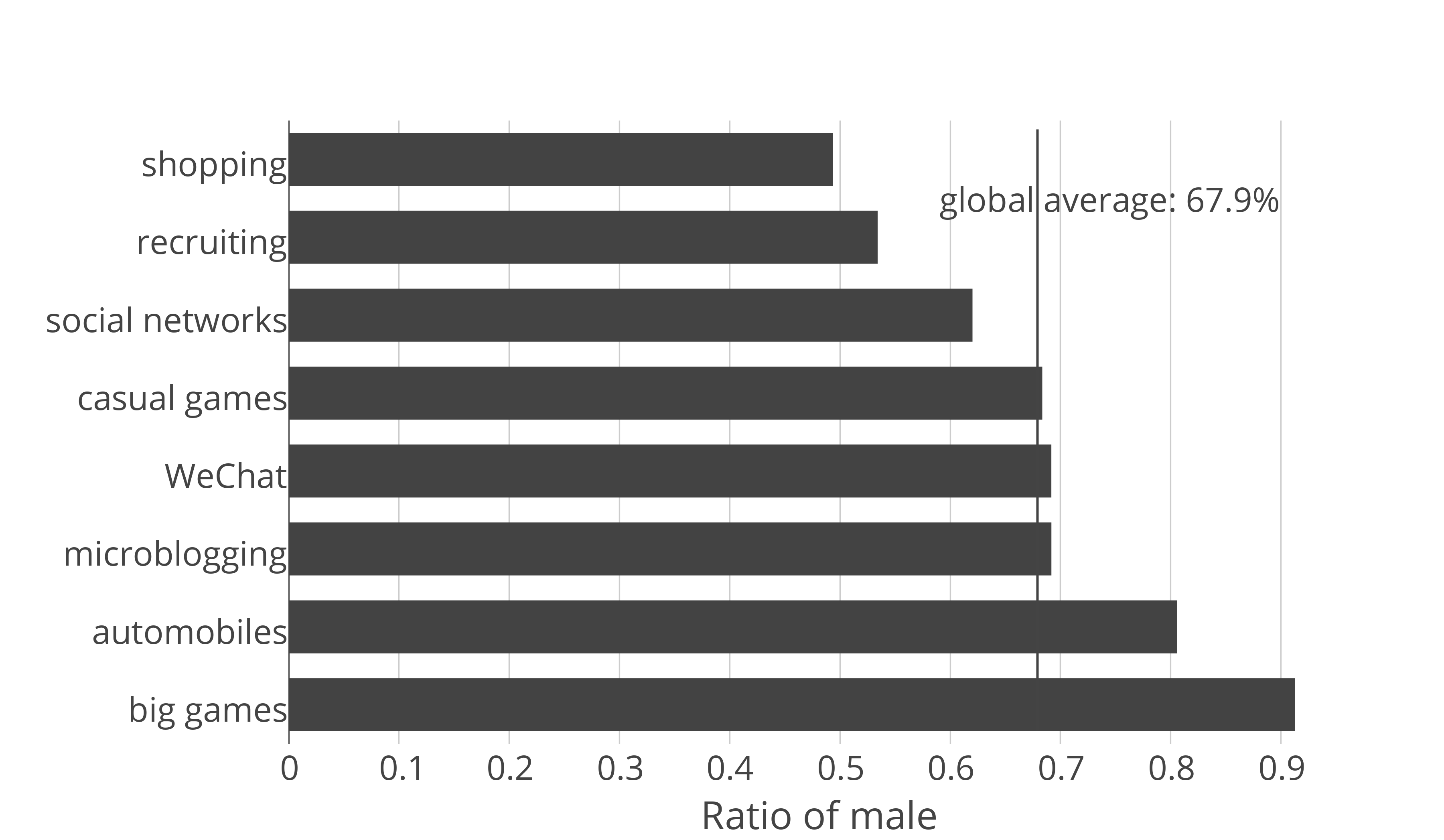}
\caption{Percentage of males in each cluster based on domains.}
\label{fig_domain_disGender}
\end{figure}

Fig. \ref{fig_domain_disGender} shows the percentage of males in each cluster. Note that the overall percentage of males is 67.9 \%. We find that there are more boys than girls visiting game websites, while there are much more girls than boys shopping online. These findings are quite intuitive in the sense that girls pay more attention on their appearance and most girls love shopping, while boys are more keen on different kinds of games, from casual web games such as \textit{Plants vs. Zombies} to big games such as \textit{DotA}.

\begin{figure}[!t]
\centering
\includegraphics[width=3.5in]{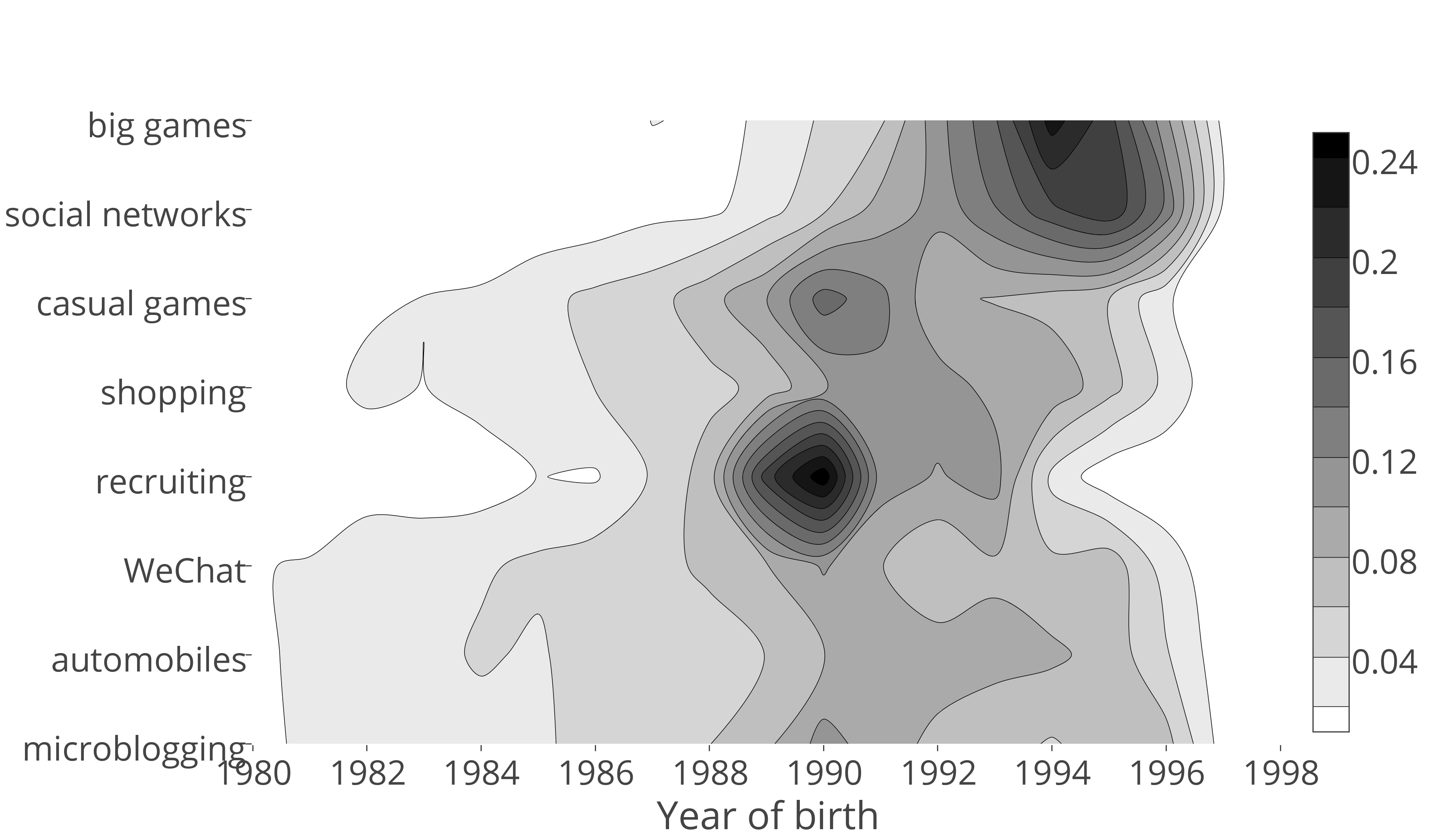}
\caption{Contour map of user birth year distributions in each cluster.}
\label{fig_domain_disAge}
\end{figure}

Fig. \ref{fig_domain_disAge} shows the user birth year distributions of each cluster on a contour map. The clusters are the same as Fig. \ref{fig_domain_cluster}.
There are two highlighted regions in this figure. The upper one shows that in the clusters of games and social networks, there are more undergraduate students. The lower one shows that in the cluster preferring recruiting websites, there are more graduate students who are likely to be searching for jobs for themselves. The results are quite straightforward that young students prefer spending more time on entertainment, while graduate students would like to pay more attention on job hunting and their future direction. Combining with Fig. \ref{fig_domain_disGender}, these findings warn the university to supervise the male undergraduates and avoid them from being addicted to the games and abandoning their studies. For graduate students, more recruitment information can be targeted to them.

\begin{figure}[!t]
\centering
\includegraphics[width=3.5in]{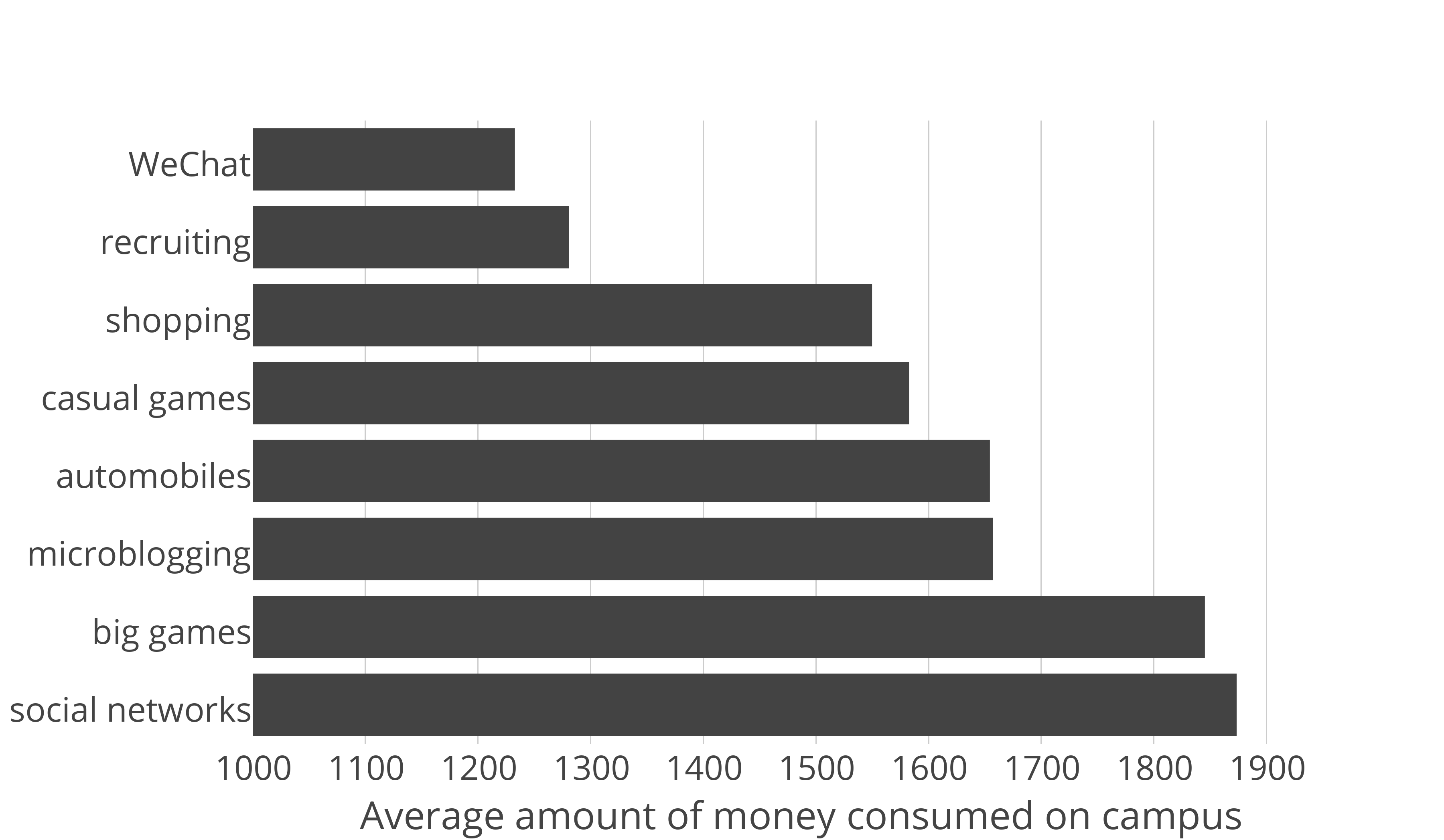}
\caption{Distribution of average campus card consumption in five months in each cluster.}
\label{fig_domain_disAmount}
\end{figure}

Fig. \ref{fig_domain_disAmount} shows the distribution of the average amount of money consumed on the campus in five months, by users of each cluster. The unit of the amount is RMB. Comparing the statistics with the age distribution in Fig. \ref{fig_domain_disAge}, we can find that undergraduate students spend more money inside the campus than graduate students. As far as we know, students in this university are divided into two campus sites. The undergraduate students are living on the campus in the suburb of the city with few business areas and restaurants. On the contrast, the graduate students are living in the urban campus site. The restaurants and supermarkets are attractive for them to spend more money outside the campus, while the undergraduate students barely have chances. Reasons may also be that the graduate students are more likely to get married and spend less time on campus, which will cause the decrease in the consumption. Therefore, to increase the turnover of the university, administrators may pay efforts to improve the attractiveness of campus dining halls, especially for professors and graduate students.

By changing the number of clusters $K$, other interesting phenomenons come out. For example, when $K=6$, the clustering results show that users holding the same brand of mobile devices, Apple and MI, are clustered into the same group. The reason may be that different brands of smart phones may tend to visit the APP domains that are within its own ecosystem. When combining with their gender, it is obvious that males prefer MI, while females prefer Apple. Therefore, when an update for either of the brands releases, the administrators for the campus networks can deploy cache for Apple updates at girls' dorms and cache for MI updates at boys' dorms.

\section{Conclusion}
\label{sec_conclusion}
In this paper, we propose a method to analyze users' internet browsing behaviour and preferences. The user behaviour modeling is based on an analogy between user behaviour profiling and topic modeling in document classification. Methods including TF-IDF and LSA are part of our process. With K-Means++ algorithm, we are able to extract hidden browsing habits by partitioning similar users into groups.

Combined with additional demographic information such as age and gender of users, the clustering results reveal more insights on user interests, which indicates the effectiveness of our proposed algorithm. The browsing habits of users vary with age and the differences between males and females are obvious. This helps us to give advice for campus management, including student management, school development and network optimization.

With larger scaled dataset, e.g., the data in a district or even a city, we may get more general user behaviour and these findings can be used in many aspects, such as commercial recommendation and network operation management. Especially, user behaviour may help with resource allocation and energy saving problems.


\section*{Acknowledgment}
This work is sponsored in part by the National Basic Research Program of China (No. 2012CB316001), and the Nature Science Foundation of China (No. 61201191 and 61401250), the Creative Research Groups of NSFC (No. 61321061), the Sino-Finnish Joint Research Program of NSFC (No. 61461136004), and Hitachi R\&D Headquarter. We also want to express our sincere appreciation for Shanghai Jiaotong University to provide all the datasets used in this work.



%
%
%

\IEEEtriggeratref{7}

\bibliographystyle{IEEEtran}
\bibliography{ref_APCC}

\end{document}